\documentclass[superscriptaddress,prl,twocolumn,amsmath,amssymb]{revtex4}
\pdfoutput=1
\usepackage{graphicx}
\usepackage{dcolumn}
\usepackage{bm}
\usepackage{subfigure}
\usepackage[final]{pdfpages}
\begin{document}

\newcommand{\tr}{\operatorname{Tr}}
\newcommand{\ket}[1]{\left | #1 \right \rangle}
\newcommand{\bra}[1]{\left \langle #1 \right |}
\newcommand{\proj}[1]{\ket{#1}\!\!\bra{#1}}

\title{Multimode quantum interference of photons in multiport integrated devices}

\author{Alberto Peruzzo}
\author{Anthony Laing}
\author{Alberto Politi}
\affiliation{Centre for Quantum Photonics, H. H. Wills Physics Laboratory \& Department of Electrical and Electronic Engineering, University of Bristol, Merchant Venturers Building, Woodland Road, Bristol, BS8 1UB, UK}
\author{Terry Rudolph}
\affiliation{Institute for Mathematical Sciences, Imperial College London, London SW7 2BW, UK}
\author{Jeremy L. O'Brien}
\email{Jeremy.OBrien@bristol.ac.uk}
\affiliation{Centre for Quantum Photonics, H. H. Wills Physics Laboratory \& Department of Electrical and Electronic Engineering, University of Bristol, Merchant Venturers Building, Woodland Road, Bristol, BS8 1UB, UK}

\begin{abstract}
We report the first demonstration of quantum interference in multimode interference (MMI) devices and a new complete characterization technique that can be applied to any photonic device that removes the need for phase stable measurements. MMI devices provide a compact and robust realization of N$\times$M optical circuits, which will dramatically reduce the complexity and increase the functionality of future generations of quantum photonic circuits.
\end{abstract}

\maketitle

Photonics is a leading approach to realizing future quantum technologies that promise secure communication \cite{gi-nphot-1-165}, tremendous computational power \cite{la-nat-464-45} and the ultimate precision measurements \cite{gi-sci-306-1330}. Optical waveguide circuits on silicon chips have demonstrated high levels of miniaturisation and performance \cite{po-sci-320-646,ma-nphot-3-346,po-sci-325-1221,laing-2010}, however, the construction of large-scale multi-port devices from $2\times2$ directional couplers makes them unwieldy and sensitive to wavelength. Multimode interference (MMI) devices promise a straightforward implementation of robust multi-port circuits, however, multi-mode operation could prevent or perturb quantum interference, which to date has not been demonstrated in this architecture. 
Here we show that quantum interference can be realised in MMI devices, observing a quantum interference visibility of $V=95.6 \pm 0.9\%$ in a $2\times2$ MMI coupler. 
We further demonstrate operation of a $4\times 4$ port MMI device with photon pairs, which exhibits complex quantum interference behaviour. 
We have developed a new technique to fully characterise such multi-port devices based on measuring quantum interference for all possible two photon input and output combinations. This approach removes the need for phase sensitive measurements and may find applications for a wide range of photonic devices. 
These results show that MMI devices can operate in the quantum regime with high fidelity and promise substantial simplification and concatenation of photonic quantum circuits.

\begin{figure}[t]
\centering
	{\includegraphics[width=\columnwidth]{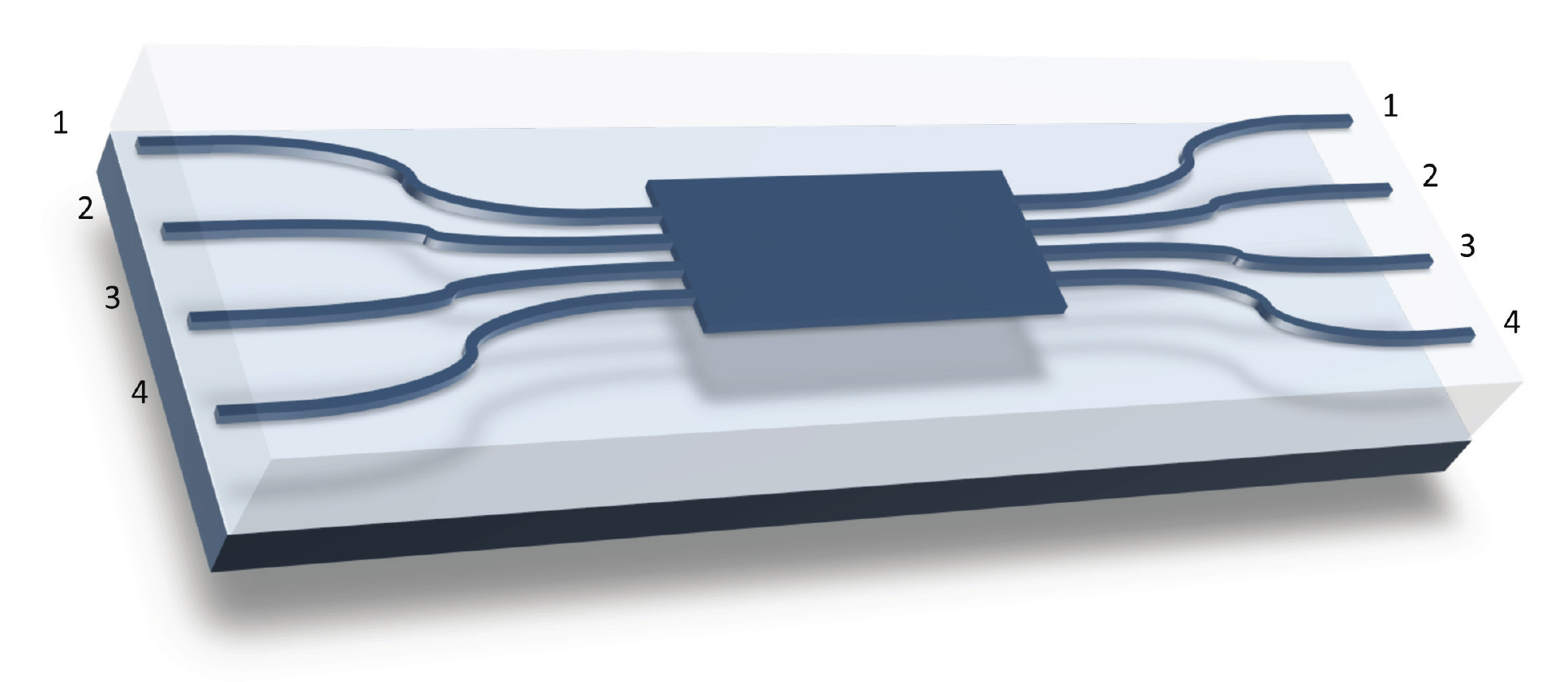}
	\includegraphics[width=\columnwidth]{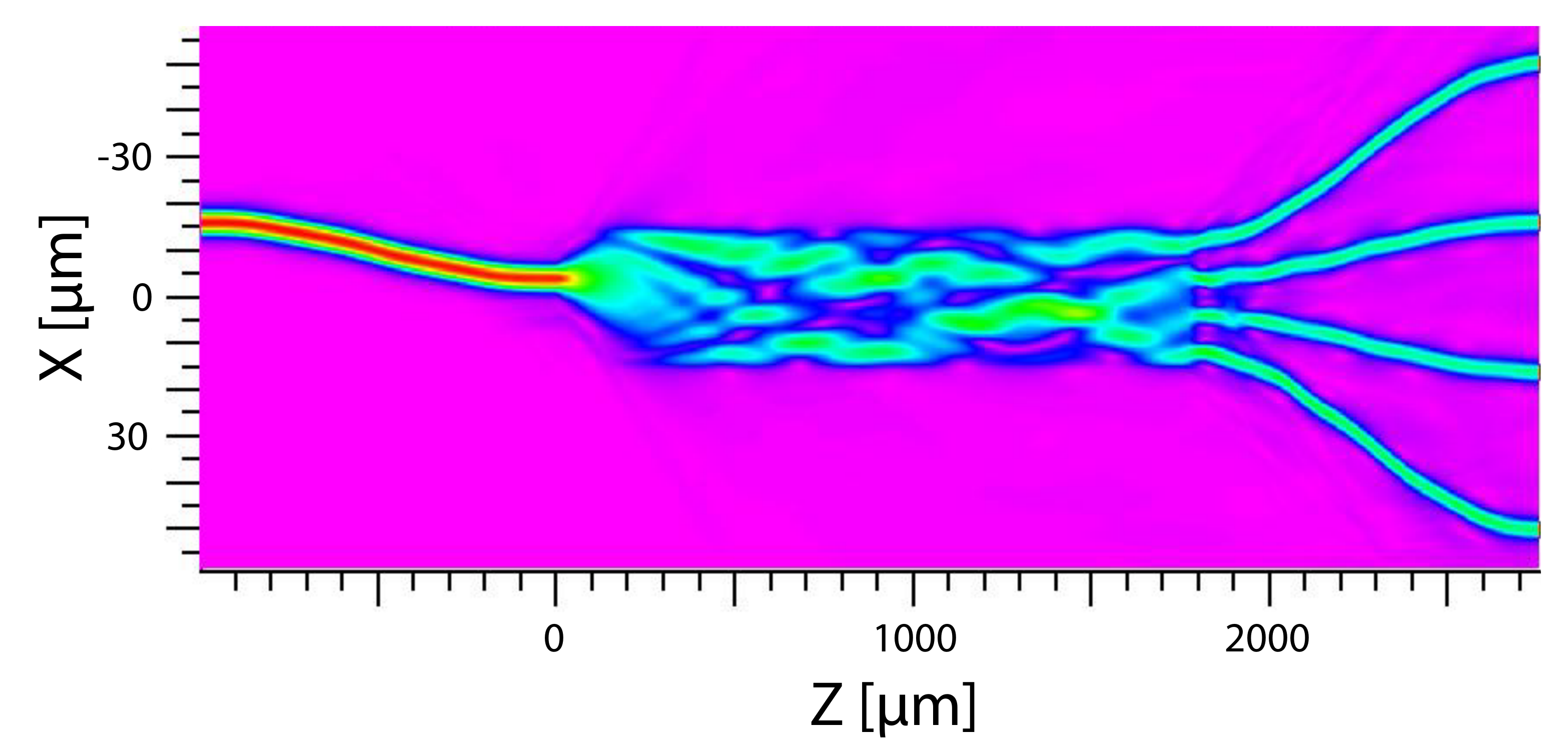}
	}	

\caption{Multimode interference devices. (a), Schematic representation of a 4x4 MMI integrated chip. (b) Simulation of classical light propagation in the 
device shown schematically in (a). Light is launched into input waveguide 2 and multimode interference in the central region results in equal intensity in each of the four output waveguides, via self imaging. Analogous behaviour is observed for injection of light in each of the other input waveguides.}
\label{schem}
\end{figure}

\begin{figure*}[t]
\centering
	\includegraphics[width=0.8\textwidth]{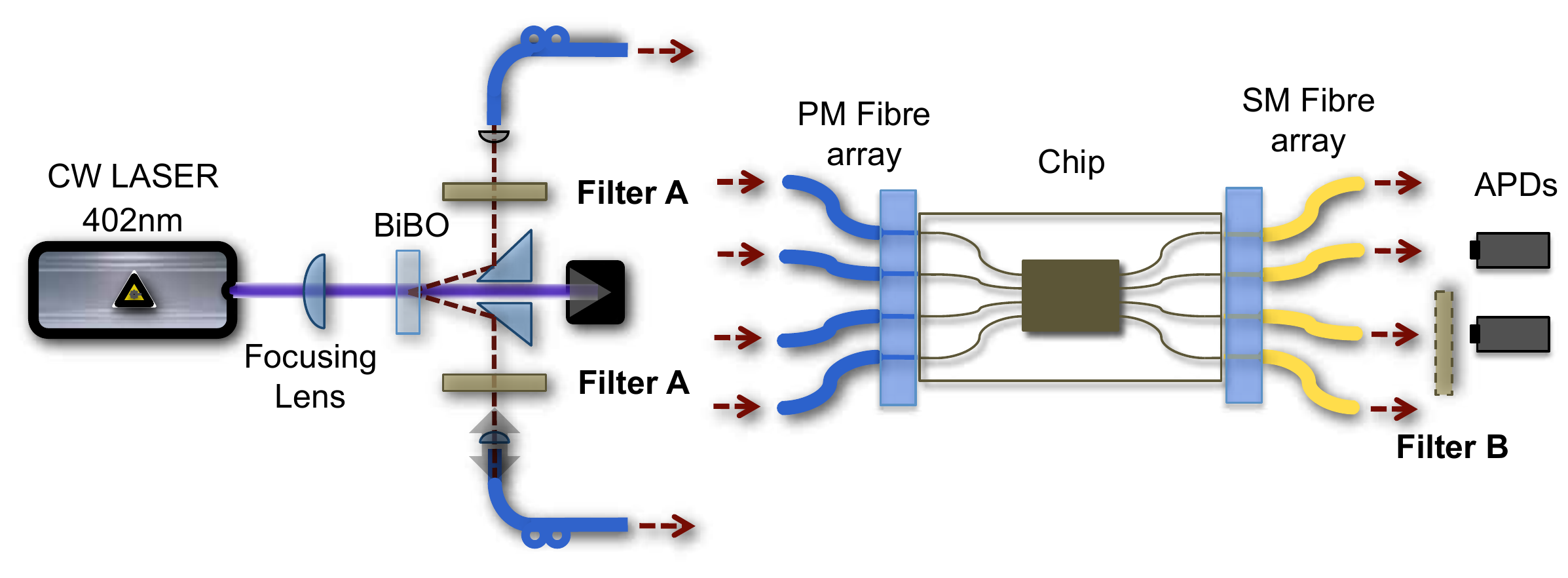}	
\caption{Experimental setup for two photon quantum interference measurements in MMI devices. For the 2$\times$2 MMI splitter filters A were 2 nm FWHM; a filter B of 0.5 nm FWFM was added to increase the coherence length. For the 4$\times$4 MMI splitter two 0.5 nm interference filters were used as filter A and no filter B was present.}
\label{exp}
\end{figure*}

Quantum technologies aim to harness superposition and entanglement to enhance communication security, provide exponential computational advantage for particular tasks \cite{de-prsla-400-97}, including factoring \cite{sh-conf-94-124}, database search \cite{gr-prl-97-325} and simulation of important quantum systems \cite{fe-ijtp-82-467}, and reach the ultimate limits of precision in measurement \cite{gi-sci-306-1330} and lithography \cite{bo-prl-85-2733}. Photons are an appealing information carrier for their inherently low-noise, high-speed transmission and the fact that entangling interactions between photons can be achieved using only linear optical circuits \cite{ob-nphot-3-687,gi-nphot-1-165,kn-nat-409-46,ob-sci-318-1567} or mediated by atom-like systems \cite{tu-prl-75-4710,de-pra-76-052312}. A photonics approach to these technologies requires complex, multi-port quantum circuits---essentially multi-path, multi-photon interferometers---that exhibit high fidelity quantum  
interference. Circuits fabricated from $2\times 2$ directional couplers have demonstrated high performance \cite{po-sci-320-646,ma-nphot-3-346,po-sci-325-1221,laing-2010,ma-oe-17-12546}, however, construction of more sophisticated multi-port circuits would require their decomposition into a very large number of $2\times 2$ directional couplers (or X-couplers \cite{sm-oe-17-13516}). For example, an arbitrary N$\times$N mode unitary \cite{re-prl-73-58} would require a sequence of $O(n^2)$ individual $2\times 2$ directional couplers.

MMI devices are based on the self imaging principle, by which an input field profile is reproduced in single or multiple images at periodic intervals along the propagation direction of a multi-mode waveguide \cite{br-josa-63-416, ul-nro6-253}. The effect is based on the propagation properties of a guide with a large number of lateral modes. MMI devices allow the design of N$\times$M splitters with superior performances, excellent tolerance to polarization and wavelength variations and relaxed fabrication requirements compared to the other main beam splitting technology, the directional couplers. Consequently MMI couplers have found applications in a broad range of photonic systems \cite{so-lt-13-615} including phase diversity networks, light switching and modulators, in laser architectures and for optical sensing applications.
In the context of photonic quantum circuits they promise to dramatically reduce the complexity of such circuits, including for example those required to generate maximally entangled path or `NOON' states \cite{pr-pra-68-052315}, W states \cite{mi-pra-70-052308} and the implementation of N$\times$N unitaries \cite{re-prl-73-58}. 

In contrast to directional couplers, the self-imaging effect in MMIs allows the flexibility to directly realize symmetric N$\times$N multi-port devices with several input and output ports. Multi-port circuits are particularly promising for quantum optics and information purposes and fundamental experiments have been conducted to study the behaviour of non-classical interference of single photons in bulk optics \cite{ma-apb-60-s111} and fibre \cite{we-pra-54-893} circuits. 
However, their performance is limited by stability and control of the splitting ratios. The implementation of multi-port splitters in MMI devices should allow higher performances due to the monolithic and scalable architecture, with the caveat that quantum interference has not been observed in such devices: multimode properties (see Fig.~\ref{schem}(b)) might at first seem undesirable in a quantum circuit since they typically perturb or destroy quantum interference. 

MMI devices, including $2\times 2$ and $4\times 4$ couplers, were designed and simulated using a commercial beam propagation package (Fig.~1(b)). They were fabricated on a 4" silicon wafer, onto which a 16 $\mu$m lower cladding layer of thermally grown undoped silica was deposited, followed by a 3.5 $\mu$m core layer of silica doped with germanium and boron oxides deposited by flame hydrolysis. This core layer was patterned into 3.5 $\mu$m wide single- and 15 and 29 $\mu$m wide multi-mode waveguides via standard optical lithographic techniques and then overgrown with a 16 $\mu$m upper cladding of  phosphorus and boron doped silica with a refractive index matched to that of the lower cladding; simulations indicated single mode operation at 780 nm. The devices are composed of N single mode waveguides that serve as input and output for the multi-mode section and terminate with a separation of 250 $\mu$m at the edges of the device to allow input and output coupling with a polarization maintaining fibre array. The  $2\times 2$ MMIs are composed of two input and two output waveguides that have a separation of 11 $\mu$m at the interface of the single- and multi-mode region. The multimode region measures $1090 \times 15$ $\mu$m. In the case of the $4\times 4$ MMIs, the waveguides are separated by 8 $\mu$m at the interface to the multi-mode section, which measures $1770 \times 29$ $\mu$m.

Ideally, a balanced $2\times 2$ MMI splitter should perform the same operation as a $2\times 2$ directional coupler with a unitary matrix that describes the evolution from input to output \cite{ma-nphot-3-346}: 
\begin{equation}
M_{2\times 2}=\frac{1}{\sqrt{2}}
 \left( \begin{array}{cc}
1 & i\\
i & 1
\end{array} \right),
\end{equation}
which equally superposes the two modes (equivalent to the Hadamard operation with a phase shift \footnote{In the $2\times 2$ case, there is only one equivalence class of symmetric splitters, this corresponds to the fact that different physical implementations can have different external phase relations, but the general description of the transformation is dictated by the unitary evolution \cite{zu-pra-55-2564}.}).
However, it is not clear that the multi-mode nature of MMI devices will allow quantum operations, in particular quantum interference.

Quantum interference with two photons is a defining distinction between classical and quantum states of light and is the key phenomenon that drives photonic quantum technologies.  Quantum interference occurs when different quantum mechanical outcomes are indistinguishable.  In the case of two photons entering the two input ports of a symmetric $2\times 2$ unitary beamsplitter (one photon per input port) the outcomes of ``both photons reflected" and ``both photons transmitted" are indistinguishable.  In this case the interference is destructive so that the photons never leave in the two separate outputs, but a superposition of two photons in each output.
This behaviour is in stark contrast to the case of two classical particles which would have a probability of 1/2 to leave in separate outputs.
When the relative arrival time of the two photons is scanned, a characteristic ``HOM" dip is observed \cite{ho-prl-59-2044} because the classical probability of 1/2 holds for finite delay and the quantum probability of zero holds for zero delay. The width of this HOM dip is related to the coherence time of the photons. The visibility $V\in [0,1]$ of the dip (how close it gets to zero) is a measure of the degree of quantum interference. Any information that distinguishes the two probability amplitudes---\emph{eg.} the photons have different polarizations, frequency, bandwidth \emph{etc.}---reduces $V<1$. A key advantage of directional coupler devices is that single mode waveguides mean that no spatial mode distinguishability is possible. In contrast, MMI devices are by definition highly multi-mode in the interaction region. 

We measured quantum interference in MMI devices using single photon pairs produced in the spontaneous parametric down-conversion (SPDC) source shown schematically in Fig.~\ref{exp}: a type-I  BiBO crystal pumped with a 60 mW CW laser diode at 40{2} nm, producing 80{4} nm pairs of photons. These photons were collected into single mode polarization maintaining (PM) fibres after passing through 2 nm filters. The source was constructed in such a way that quantum interference with $V\approx 98.5 \%$ was routinely observed, confirmed with a directional coupler that has previously exhibited $V=1$ (Ref.~\onlinecite{laing-2010}) \footnote{All the visibilities quoted in this paper except those for the 4x4 MMI are corrected for accidentals \cite{laing-2010}}.

\begin{figure}[t]
\centering
	\subfigure[]{\includegraphics[width=0.4\textwidth]{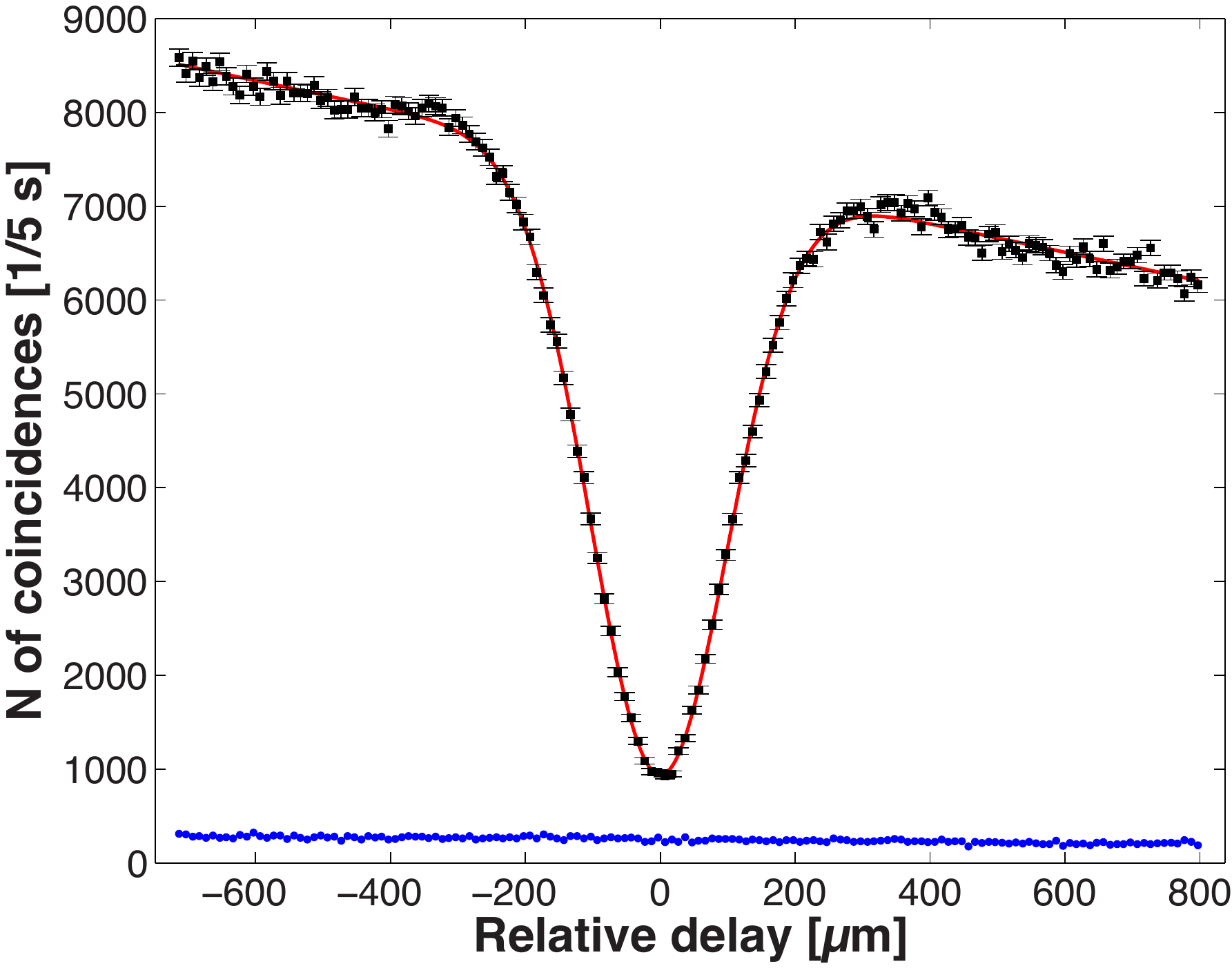} \label{fig_first_case}}
	\subfigure[]{\includegraphics[width=0.4\textwidth]{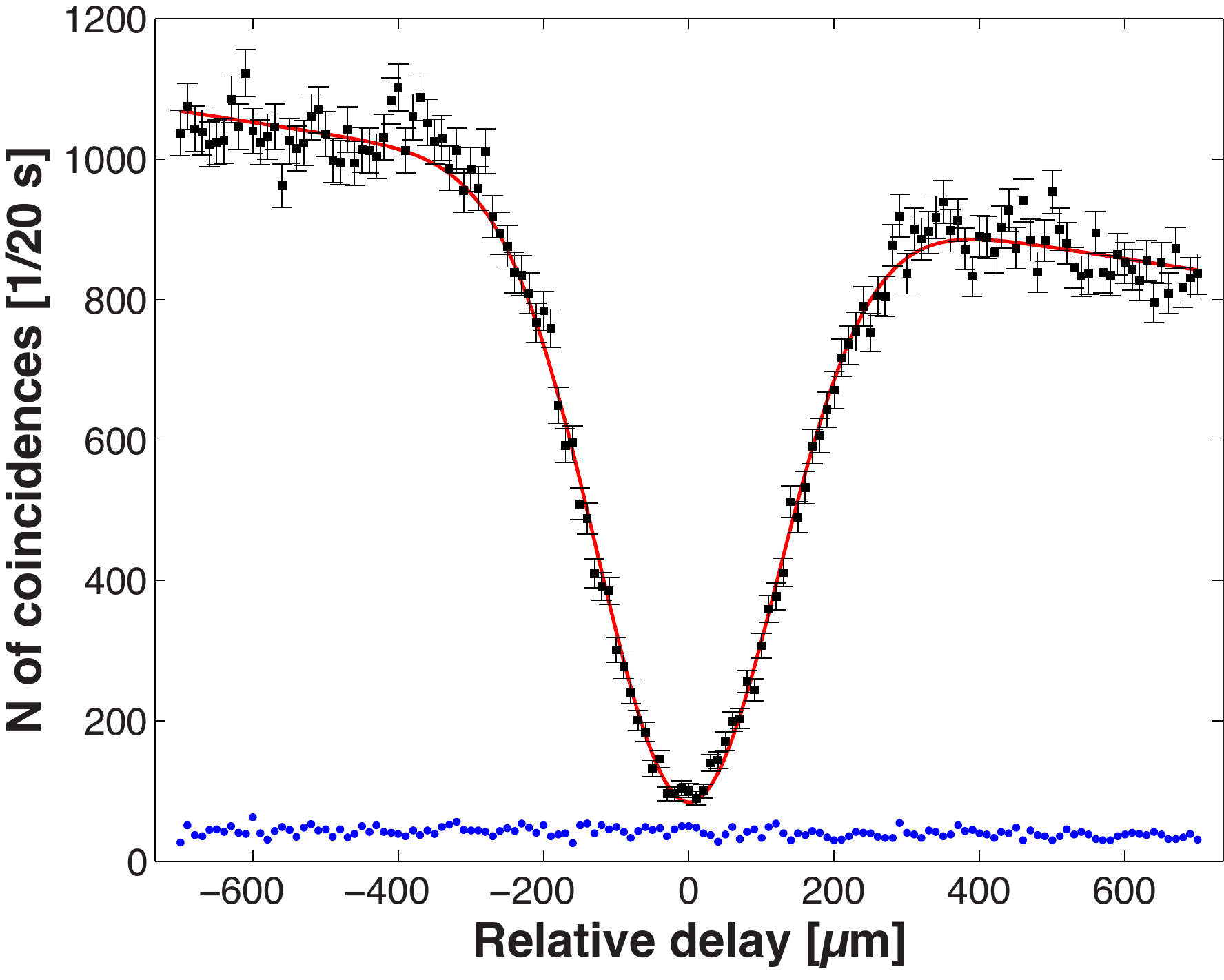} \label{fig_second_case}}
\label{fig_MMI2}
\caption{Quantum interference in a $2\times 2$ MMI coupler. (a) The measured HOM dip for 2 nm filters, corresponding to a dip FWHM of 239 $\mu m$. (b) The measured HOM dip for the same device and source, but with an additional 0.5 nm filter inserted into one output, resulting in a dip FWHM of 296 $\mu m$. The blue data show the measured rate of accidental counts.}
\end{figure}

We observed the HOM dip shown in Fig.~\ref{fig_first_case} in a $2\times 2$ MMI coupler. These data provide conclusive evidence that quantum interference does indeed occur in a MMI device (the linear slope in these data is due to decoupling of the input fibre as the timing delay is changed). However, the measured $V = 90.4 \pm{0.4} \%$ is significantly lower than the $V\approx 98.5 \%$ obtainable from the SPDC source. The reason is that the propagation in the multi-mode section of the MMI introduces some distinguishability between the photons. We experimentally ruled out spatial, spectral and polarization mismatch of the photons, implicating the temporal degree of freedom. The different modes in the multimode section of the device see different effective refractive indices, which introduces a jitter in the time of flight of the photons from the input to the output waveguides, providing ``which path" distinguishing information, and thereby reducing $V$. 

To confirm that this temporal jitter effect is the origin of the reduced visibility, we inserted a narrower 0.5 nm filter in one of the output modes between the device and the detector (as indicated in Fig.~\ref{exp}), i.e. not affecting the properties of the photon source, but simply increasing the coherence length of the photons. The additional filter acts as a quantum eraser \cite{kw-pra-45-7729}, that erases the timing information by increasing the coherence time of the photons. Under these experimental conditions, we observed the HOM dip plotted in Fig.~\ref{fig_second_case} in the same $2\times 2$ MMI device. In this case $V=95.6 \pm{0.9\%}$, confirming that timing jitter limits the visibility for the data shown in Fig.~\ref{fig_second_case} (the larger error bar is due to the lower count rate with the narrower filter). These data confirm that quantum interference occurs in MMI devices, and that the coherence length of the photons must be sufficiently long compared to the timing jitter that is introduced as a result of the different refractive indices of the MMI modes.

\begin{figure*}[t]
\centering
	\includegraphics[width=\textwidth]{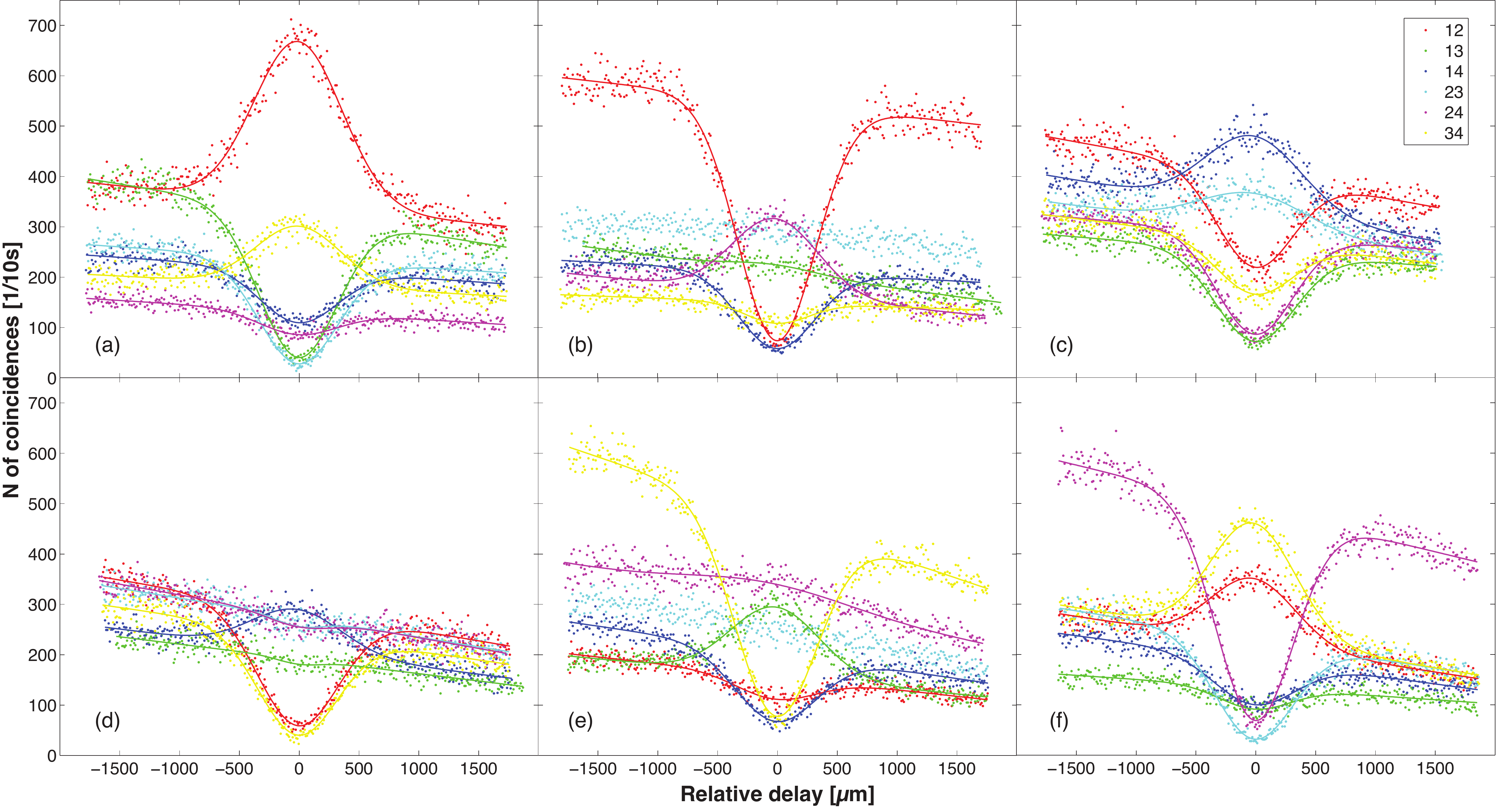}
\caption{Quantum interference in a $4\times 4$ MMI coupler. Coincidence counts of two photons at the output ports of a $4\times 4$ MMI device as the arrival time of the photons is varied. The different graphs represent the six possible input states in the splitter: (a) $\ket{11}_{12}$, (b) $\ket{11}_{13}$, (c) $\ket{11}_{14}$, (d) $\ket{11}_{23}$, (e) $\ket{11}_{24}$, (f) $\ket{11}_{34}$. The FWHM of $\sim 800 ~ \mu m$ is as expected for the 0.5 nm interference filters used.}
\label{fig_4x4MMIplots}
\end{figure*}

Interestingly, the description of multi-port splitters grows in complexity with N. For N $\leq3$ all symmetric N $\times$ N splitters can be described by one equivalence class, since the requirement on the conservation of energy defines the matrix to within external phases on the input and outputs. However, when N $\geq4$ there exists an infinite number of distinct equivalence classes \cite{be-jmp-15-1677} and internal free phases are independent of the conservation of energy \cite{pr-pra-68-052315}. The transition matrix that describes an ideal symmetric $4\times 4$ MMI splitter is:
\begin{equation}
M_{4\times 4,\theta}= \frac{1}{2}
 \left( \begin{array}{cccc}
1 & 1 & 1 & 1 \\
1 & e^{i \theta} & -1 & -e^{i \theta} \\
1 & -1 & 1 & -1\\
1 & -e^{i \theta}  & -1 & e^{i \theta}
\end{array} \right)
\label{eq_M8}
\end{equation}
where $\theta$ is the free internal phase. In general, two different physical implementations would correspond to a different equivalence class, and a different value of $\theta$.

In the case of a MMI splitter the value of the internal phase is, in principle, dictated by the self-imaging condition \cite{so-lt-13-615}. However, the presence of fabrication imperfections in the device would drive the multi-mode section away from exact self-imaging, and the relation between the optical phases would deviate from the expected value. Moreover, the presence of unavoidable losses in the MMI coupler corresponds to the presence of additional optical modes in the transition matrix of the splitter, thus making the reduced ${4\times 4}$ transition matrix $M_{}$ more complicated than Eq.~\ref{eq_M8}. In principle, $M_{}$ could be reconstructed using a number of phase sensitive measurements. Such measurements are, however, difficult in practice, due to the need to maintain sub-wavelength stability in interferometers consisting of waveguide and fibre and/or free space paths. We have developed a technique to overcome this using only intensity measurements and two photon quantum measurements, but no phase sensitive measurements, as described below. 

\begin{figure*}[t!]
\centering
	\includegraphics[width=0.85\textwidth]{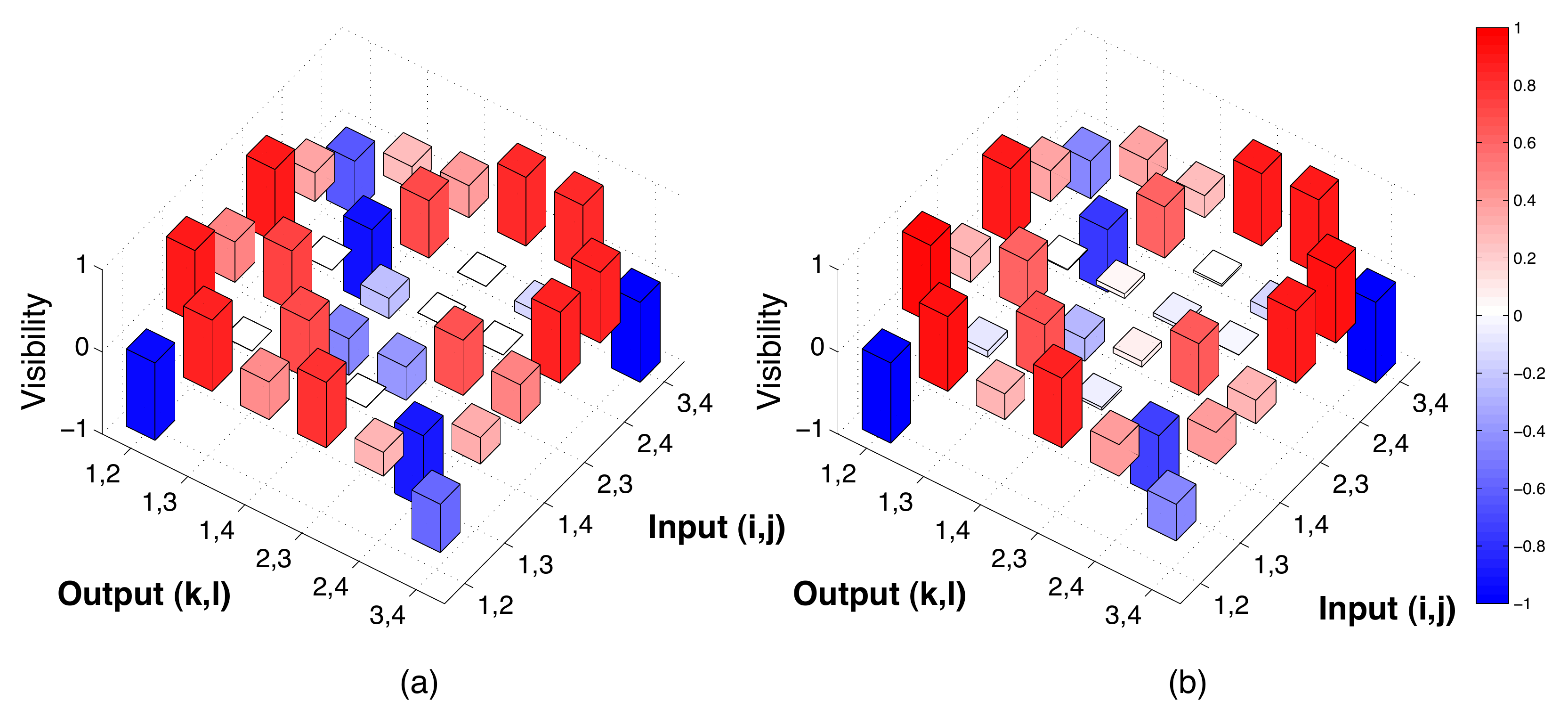}
	\vspace{-.4cm}
\caption{Quantum interference matrix of a $4\times 4$ MMI coupler. Measured (a) and reconstructed (b) visibility matrices $V_{ijkl}$ of the non-classical interference between two photons injected in input waveguides $i$ and $j$ and detected in output waveguides $k$ and $l$ of a 4$\times$4 MMI splitter. Positive visibilities correspond to a HOM-like dip, negative to peaks.}
\label{fig_4x4MMIDipsMeasured}
\end{figure*}

In contrast to the $2\times 2$ MMI device in which quantum interference is destructive, the $4\times 4$ MMI device interference between indistinguishable outcomes can be constructive for some of the 36 possible input and output combinations \cite{ma-apb-60-s111}. 
We characterised the quantum operation of a $4\times 4$ MMI splitter by inputing the state $\ket{11}_{ij}$---\emph{i.e} a single photon in each input waveguide $i$ and $j$. We considered all 
six combinations of two photons in four inputs, where $i\neq j$ (we did not consider the case of both photons in the same input since this does not give rise to quantum interference---\emph{i.e.} such measurements provide no more information than bright intensity measurements do). 
As in the case of the HOM dip, quantum interference is revealed in the correlations in the output probability distribution. The probability to detect one photon in each output $\textcolor{black}{k}$ and $l$, when two indistinguishable photons are injected into inputs $i$ and $\textcolor{black}{j}$ is given by:
\begin{equation}
Q_{ij}^{kl} = \frac{1}{1+\delta_{ij}}|M_{ik}M_{jl}+M_{il}M_{jk}|^2,
\label{eq_Pq}
\end{equation}
where $\delta_{ij}$ is Kronecker's delta and $M_{ij}$ is the element of the transition matrix. In the case of two distinguishable photons (equivalent to the classical analogue) 
the probability is given by
\begin{equation}
C_{ij}^{kl} = |M_{ik}M_{jl}|^2+|M_{il}M_{jk}|^2,
\label{eq_Pc}
\end{equation}
and there is no interference between the two terms.  
Measurement of the generalized non-classical interference between two photons enables the reconstruction of the matrix $M_{}$ via measurement of the detection probabilities $Q_{ij}^{kl}$ and $C_{ij}^{kl}$, as described below.

Figure \ref{fig_4x4MMIplots} shows the rate of detecting two photons at the six possible output pairs of waveguides, for each of the six different input states of a $4\times 4$ MMI splitter, as a function of the relative arrival time of the photons.  
For some input-output combinations, these data exhibit clear interference dips, analogous to the HOM dip. However, for other combinations there are peaks, and essentially straight curves.
These behaviours are the result of the phase values of the matrix $M_{}$ and are conveniently summarised by plotting the visibility of the non-classical peak or dip, given by:
\begin{equation}
V_{ijkl}=\frac{ C_{ij}^{kl}-Q_{ij}^{kl}}{C_{ij}^{kl}},
\end{equation}
where positive values indicate a dip and negative values a peak. Fig. \ref{fig_4x4MMIDipsMeasured}(a) shows the 6$\times$6 matrix of measured visibilities $V^m$ obtained from the data of Fig. \ref{fig_4x4MMIplots}. 

We have developed a technique that uses only the values $V_{ijkl}$ and the classical intensity ratios $|M_{ik}|$ to reconstruct the (reduced) transition matrix that describes the MMI device, assuming linearity of the device. To do this, we numerically search for a matrix $M^{r}$ that minimizes the RMS distance  
between the experimentally measured $V^m$ and the reconstructed $V^{r}$ that corresponds to $M^{r}$. 
The classical intensity ratios $|M_{ik}|$ were measured using a CCD camera at the output of the integrated chip. 
The numerical optimization produces\footnote{Note that since the elements of each row are normalized to 1 and the
device includes some loss, which is considered explicitly via the addition of more modes, the sum of each column
can be greater $>$1.}

\begin{equation*}
M^{r}=
 \left( \begin{array}{cccc}
  0.72 & 0.49 & 0.43 & 0.24 \\
  0.62 & 0.37e^{i0.06}  & -0.52e^{-i0.06} & -0.18e^{-i0.41} \\
  0.56 & -0.60e^{-i0.04} & 0.03e^{-i0.33} & -0.42e^{i0.22} \\
  0.35 & -0.21e^{-i0.47} & -0.42e^{i0.22} & 0.47e^{i0.41}
\end{array} \right).
\label{eq_M8_extended}
\end{equation*}
  
The quality of the reconstructed transition matrix $M^{r}$ is confirmed by comparing the measured $V^m$ and calculated  $V^{r}$ matrices in Fig. \ref{fig_4x4MMIDipsMeasured}. Additional optimization routines were performed with different initialization parameters, to confirm that the measured transmissivities do indeed provide a good starting condition for the matrix $M^{r}$. It is interesting to note that this method to reconstruct the transition matrix can be used for any unknown linear optical element (where no decoherence or measurement is present), even in the case of losses. It is possible to calculate the reduced $n\times m$ transition matrix even in the case of large N$\times$N (N$>n,m$) networks in which only $n$ input and $m$ output modes are accessible for measurements. 

Interestingly, the computation of the photon distribution at the output of a multi-port circuit with many photons is a hard problem to solve from a computational perspective. The coincidence probability distribution is related to the permanent of the matrix that describes the multiport device \cite{sc-quant-ph-0406127}. Since the computation of the permanent of a matrix is a computationally difficult problem, the measurement of the coincidence probability in a device with many photons and many ports could represents a feasible implementation of quantum simulation that uses the bosonic nature of the photons to realize a hard computation. In general, it is possible to map the calculation of the permanent to the detection of multiphoton coincidences in the appropriate splitter. Although the form of the transition matrix that characterizes the MMI device is in principle fixed by the self-imaging condition, and no reconfigurability is possible, the data presented here on a $4\times 4$ MMI splitter is the first small-scale example of such a quantum computation \cite{aaronsonQIP2010}. 

Multi-photon inputs to multi-port devices are not only an essential ingredient of future photonic quantum technologies, but enable the study of a rich variety of quantum interference phenomena to be studied 
\cite{ca-pra-62-013809}.  The increasing need for more complex photonic networks will present the problem of the characterization of such integrated circuits and their imperfections. As demonstrated here, it is possible to take advantage of the properties of quantum states of light to reconstruct the behaviour of a photonic network, without the need of complex phase stable measurements. 
In the case of quantum networks, MMI devices promise to dramatically simplify and miniaturize the implementation of quantum circuits, thanks to the possibility of performing complex multi-mode evolution of many photons in a single compact device.

\vspace{2pt}

We thank J. C. F. Matthews and J. G. Rarity for helpful discussions. This work was supported by EPSRC, ERC, the Leverhulme Trust, QIP IRC, IARPA and NSQI.
J.L.O'B. acknowledges a Royal Society Wolfson Merit Award.
\vspace{-12pt}

\bibliography{bib17b,bib15a}

\end{document}